\documentstyle[aps,prl,amsfonts,psfig]{revtex}
\newcommand{\beq}{\begin{equation}}
\newcommand{\eeq}{\end{equation}}
\begin{document}

\title{\bf Chaos in effective classical and quantum dynamics}

\author{Lapo Casetti\cite{lapo}, Raoul Gatto\cite{gatto}
and Michele Modugno\cite{michele}}
\address{D\'epartement de Physique Th\'eorique, Universit\'e de Gen\`eve,
24 Quai Ernest-Ansermet, CH-1211 Gen\`eve, Switzerland}

\date {\today}
\maketitle


\begin{abstract}
We investigate the dynamics of classical and quantum $N$-component ${\phi}^4$ 
oscillators in the presence of an external field. In the large $N$ limit 
the effective dynamics is described by two-degree-of-freedom classical
Hamiltonian systems. In the classical model we observe 
chaotic orbits for
any value of the external field, while in the quantum case chaos is strongly
suppressed. A simple explanation
of this behaviour is found in the change in the structure of the orbits
induced by quantum corrections. Consistently with Heisenberg's principle, 
quantum fluctuations are forced away
from zero, removing in the effective quantum dynamics a
hyperbolic fixed point that is a major source of chaos in the classical model.
\end{abstract}


The study of the quantum mechanics of those systems whose classical
counterpart exhibits chaotic dynamics has attracted a lot of
interest in recent years,
and is an open and rapidly evolving field \cite{LesHouches,Casati}.
Chaos does not exist in the linear evolution of the quantum state
vector, hence different approaches to identify quantum features
that correspond to classical chaos 
have been developed, ranging from the
application of random matrix theory to the
statistical analysis of energy spectra \cite{Bohigas} and  to
various semiclassical approximations \cite{semiclassical}.
Dynamical chaos in the actual quantum evolution may show up 
in mean-field approaches \cite{presilla}, or using
Bohm's formulation of quantum mechanics to define quantum
trajectories and quantum Lyapunov exponents \cite{Guglielmo}.
Moreover, at a semiclassical level, the dynamics of quantum 
expectation values can be chaotic \cite{semiquantum,elze,exact}.
 
In this Letter we consider the effective dynamics of quantum
expectation values as obtained in the large $N$ limit.
The purpose of the present work is twofold. First, we want to compare the
quantum effective dynamics of a model system
with the classical effective dynamics of the same system
at the same level of approximation, in order to unambigously detect
the effect of the quantum corrections on dynamical chaos. This effect
turns out to be a strong suppression of chaos with respect to the 
classical case. 
Second, we want to show
that in our model the suppression of chaos in the quantum dynamics
has a clear physical origin in the fact that quantum fluctuations must
be nonvanishing, i.e., in the Heisenberg principle.

As a model system we consider a $N$-component $\phi^4$ field 
theory in $d+1$ 
space-time dimensions, in the presence of an external
field $B$. We shall limit ourselves to the case $d=0$,
which allows us to offer a simple intuitive explanation of the effectiveness
of quantum corrections in suppressing chaos.
The Lagrangian that we consider is ($\alpha = 1,\dots, N$)
\beq
{\cal L} =  \frac{1}{2}\dot{\phi}^{\alpha}\dot{\phi}_{\alpha}
+ \frac{1}{2}\mu^2 \phi^{\alpha}\phi_{\alpha}
- \frac{\lambda}{8 N} (\phi^{\alpha}\phi_{\alpha})^2
+ B \frac{1}{N}\sum_{\alpha}\phi_{\alpha}~.
\label{ham}
\eeq
We perform a $1/N$ expansion, keeping only the leading order term
in both the classical and quantum case. In the former case we start by writing
\beq
\phi_{\alpha} = \frac{1}{N}\sum_{\beta}\phi_{\beta} + \delta \phi_{\alpha}
\equiv \varphi + \delta \phi_{\alpha}
\label{phi}
\eeq
and we approximate the quadratic fluctuations by considering all of them
equivalent in the large $N$ limit
\beq
\delta \phi_{\alpha}^2 \simeq \xi^2~.
\label{delta}
\eeq
By inserting Eqs.\ (\ref{phi}) and (\ref{delta}) in Eq.\ (\ref{ham})
 it turns out that the dynamics of
 the mean field $\varphi$ and of its root mean square fluctuation $\xi$
is governed by the following effective Hamiltonian
\beq
{\cal H} = \frac{1}{2}\left(\pi^2 + \eta^2 \right) + \frac{\lambda}{8}
\left(\varphi^2 + \xi^2 - v_0^2 \right)^2  - B \varphi~.
\label{H_class}
\eeq
where the two canonically conjugated pairs of variables are
$\varphi,\pi$ and $\xi,\eta$, and $v_0 \equiv \sqrt{2\mu^2 /\lambda}$
is the minimum of the potential energy in Eq.\ (\ref{ham}) for $B=0$.

In the quantum case we consider the time evolution of the expectation
value $\phi\equiv\sum_{\alpha}\langle\phi_{\alpha}\rangle/N$ from a given
initial quantum state.
This initial value quantum problem can be formulated by using
the ``closed time path'' functional formalism \cite{ctp}.
The application
of this formalism to the present case was developed by Cooper {\em et al.}
\cite{cooper},
who showed that the evolution equations in the large $N$ limit
are (classical) Hamilton's equations for the effective Hamiltonian
(we keep the dependence on the external source $B$),
\beq
{\cal H} = \frac{1}{2}\left(\pi^2 + \eta^2 \right) + \frac{\lambda}{8}
\left(\varphi^2 + \xi^2 - v_0^2 \right)^2 +
\frac{\hbar^2 \sigma^2}{8 \xi^2} - B \varphi~,
\label{H}
\eeq
where $\sigma\equiv 2 n + 1$, $n = \langle a^{\dagger}a \rangle$
being the expectation value, in the initial state,
of the particle number operator for a single
oscillator, and $\xi$ the expectation value
of the r.m.s. fluctuation of the fields, in close analogy to the 
classical case.
In Ref. \cite{cooper} it was also shown that ${\cal H}$ in
Eq.\ (\ref{H}) is just the
expectation value of the full quantum Hamiltonian in
a general mixed (initial) state characterized by a Gaussian density matrix.

We notice that at this order of the expansion the correspondence
between the classical and the quantum case is very strict,
since both are described in terms of corresponding degrees of freedom,
the mean field $\varphi$ and its r.m.s. fluctuations $\xi$,
and the classical effective Hamiltonian is just the quantum one with
$\hbar = 0$. This shows that the approximation (\ref{delta})
is equivalent to retaining only the classical contribution to the fluctuations.
The quantum correction to the Hamiltonian (\ref{H}) keeps the fluctuation
$\xi$ away from zero, consistently with Heisenberg's uncertainty principle.
In the following we will always refer to the Hamiltonian (\ref{H}),
distinguishing between the classical and the quantum case
according to the value of $\hbar$.

In the case of a vanishing external field ($B = 0$) the Hamiltonian
(\ref{H}) is integrable in both the classical ($\hbar = 0$) and quantum
case ($\hbar \not = 0$). The integrals of motion are the
total energy $E$ and the function
\beq
I = \left(\varphi\eta -\xi\pi \right)^2 + \frac{\hbar^2 \sigma^2}{4 \xi^2}
\varphi^2~.
\label{I}
\eeq

The conservation of $I$ in Eq.\ (\ref{I}) is due to the fact
that the Hamiltonian (\ref{H}) can be seen as the  Hamiltonian,
in cylindrical coordinates, of a particle in three dimensions
moving in a central potential,
$\varphi$ being the azimuthal coordinate and $\xi$ the radial one.
The term ${\hbar^2 \sigma^2}/{4 \xi^2}\varphi^2$ is indeed the centrifugal
barrier term  \cite{cooper} which shows up passing 
from cartesian to cylindrical
variables. From conservation of angular momentum
it follows that $I \equiv L^2 - L^2_z$ is conserved.

Switching on the external field ($B\not = 0$),
the two systems become nonintegrable. They
have two integrable limits. These limits are:
harmonic oscillators as $E \to 0$; and
the integrable Hamiltonian with $B = 0$ as $E \to \infty$. In fact,
as the perturbation is linear in $\varphi$, it will become negligible
with respect to the other terms in ${\cal H}$ for sufficiently large $E$.
Hence, we expect that chaotic orbits may show up in an intermediate
energy range, whose width will depend on $B$.
We have studied the dynamics by numerically integrating
the canonical equations of motion
derived from the Hamiltonian (\ref{H}) using a bilateral symplectic
algorithm \cite{Casetti}.
The values of the parameters were fixed by working in natural units
$\hbar = 1$ (in the quantum case) and by chosing
$\lambda = v_0 = \sigma=1$. The choice $\sigma=1$ corresponds to an
initial vacuum state for the number operator $a^{\dagger}a$ \cite{cooper}.

We have studied chaos from both a qualitative
and a quantitative point of view, i.e., we have calculated Poincar\'e sections
\cite{Lichtenberg} in the plane $(\xi,\eta)$
and measured the Lyapunov exponent \cite{Lyapunov} 
of every single orbit. This has allowed
us to also obtain an estimate of the relative measure $\mu$
of the chaotic regions in the phase space, defined as the ratio of the number of
trajectories whose Lyapunov exponent is positive to the total number
of trajectories, so that $0 \leq \mu \leq 1$.
For each value of the energy $E$ and of the field $B$ we have estimated
$\mu(E,B)$
from a sample of 1000 orbits picked up at random in the allowed region
on the section surface. We have considered the energy range $0 \leq E \leq 1$,
always fixing the zero of $E=0$  as the minimum energy allowed in the classical
$B=0$  case, and field intensity range $0 \leq B \leq 0.5$.

\begin{figure}
\centerline{\psfig{file=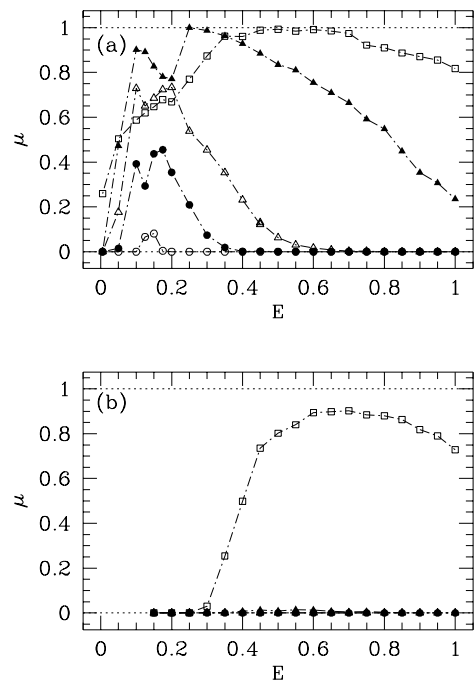,height=10cm}}
\caption{Relative measure $\mu$ of the chaotic component of phase space
vs energy $E$ at different values of the external field $B$.
Each point is an average over a sample of 1000 randomly chosen orbits.
$(a)$ classical case; $(b)$ quantum case.
Symbols in both cases: $B = 0.01$ (circles), $B = 0.05$ (solid circles),
$B = 0.1$ (triangles), $B = 0.3$ (solid triangles), $B = 0.5$ (squares).
Errorbars are of the same size as the data points.
\label{fig_mu}}
\end{figure}

\begin{figure}
\centerline{\psfig{file=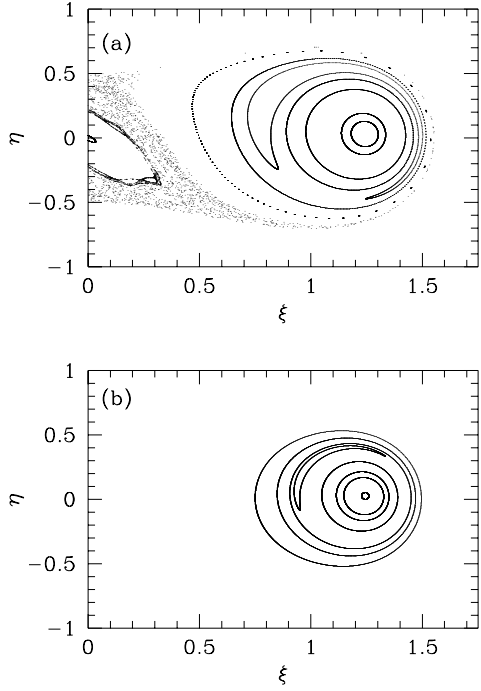,height=10cm}}
\caption{Poincar\'e section of the Hamiltonian flow defined by
Eq.\ (\protect\ref{H}) with $B = 0.05$. $(a)$ classical case,
$(b)$ quantum case; the energy is $E = 0.25$ in both cases.
\label{fig_poincare_lowb}}
\end{figure}

\begin{figure}
\centerline{\psfig{file=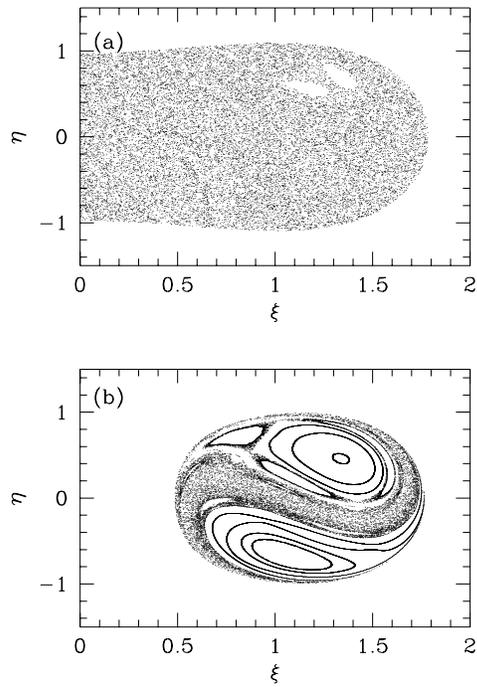,height=10cm}}
\caption{The same as in Fig. \protect\ref{fig_poincare_lowb} with
$B=0.5$ and $E = 0.6$.
\label{fig_poincare_highb}}
\end{figure}

In Fig. \ref{fig_mu} we plot $\mu(E,B)$ for the classical
and the quantum case. It is evident that, as soon as $B \not = 0$,
chaotic orbits suddenly appear in the classical case.
For small values of $B$, i.e., $B = 0.01$,
such orbits are present only in a small interval of energies
centered around $E \simeq 0.1$, then the chaotic energy interval
broadens as $B$ grows, and eventually fills the whole explored energy
range as $B$ becomes larger than $0.3$. In the quantum case the situation
is completely different: no chaotic orbits are detected for $B \leq 0.3$,
then chaos appears but at considerably larger values of $E$ as
compared to the classical case. To give an example, in
Figures \ref{fig_poincare_lowb} and
\ref{fig_poincare_highb} we show a comparison between Poincar\'e sections in
the classical and
the quantum case at two different values of the external field
(notice that the external field $B$ does not affect the shape of the
region of the plane $(\xi,\eta)$ accessible to the system,
since $B$ is coupled to $\varphi$).

\begin{figure}
\centerline{\psfig{file=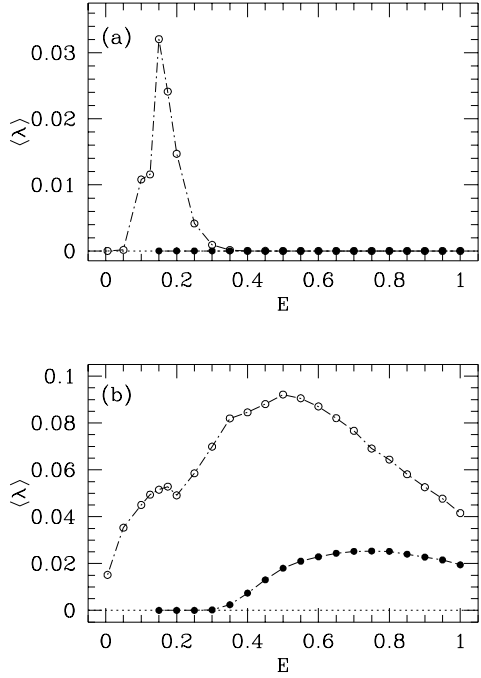,height=10cm}}
\caption{Comparison between the values of $\langle\lambda\rangle(E)$
in the classical (open circles) and quantum (solid circles) case
at $(a)$ $B = 0.05$ and $(b)$ $B=0.5$.
\label{fig_lyap}}
\end{figure}

In order to quantify the degree of chaos
at a given energy $E$ and external field $B$ we have considered also the
ensemble
average $\langle\lambda\rangle(E,B)$ of the Lyapunov exponent over the samples
of 1000 trajectories used to compute $\mu(E,B)$.
A comparison between the classical and the quantum case is reported in Fig.
\ref{fig_lyap} for the same values of $B$
as in Figs. \ref{fig_poincare_lowb} and \ref{fig_poincare_highb}.
Looking at these figure it is evident that chaos is strongly
suppressed in the quantum case with respect to the classical case.

\begin{figure}
\centerline{\psfig{file=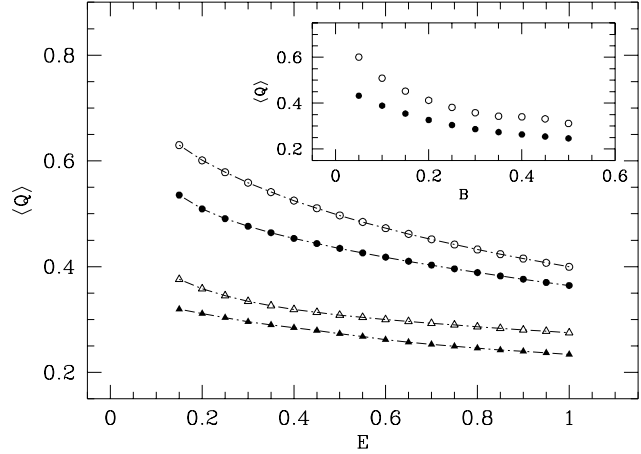,height=10cm}}
\caption{Quantum parameter $\langle Q \rangle$ (see text) vs the energy $E$ at
different values of the field $B$. Symbols as in Fig. \protect\ref{fig_mu}.
Inset: $\langle Q \rangle$ vs the field $B$ for two values of
$E$, $E=0.2$ (solid circles) and $E=0.8$ (circles).
\label{fig_Q}}
\end{figure}

As $E \to \infty$ the quantum and classical models are equivalent,
hence it is worth considering, in addition to the average measures
of chaos $\mu(E,B)$ and $\langle\lambda\rangle(E,B)$, also an average
measure of the relative importance of the quantum part of the Hamiltonian,
given by the ensemble average $\langle Q \rangle(E,B)$, where for each single
orbit $Q$ is defined as
\beq
Q = \frac{\langle V_Q \rangle_t}{\langle V \rangle_t}~,
\eeq
Here $V_Q = (\hbar^2 \sigma^2)/(8 \xi^2)$ is the quantum correction to the
potential, and $V$ is the total potential --- suitably normalized
\cite{normalization} in order
that $0 \leq Q \leq 1$ --- and $\langle \cdot \rangle_t$ stands for a time
average along the orbit. The parameter $Q(E,B)$ has a smooth dependence
on both $E$ and $B$, at variance with the other parameters (Fig.
\ref{fig_Q}). The transition
from completely ordered, to mixed (ordered + chaotic),
to almost completely chaotic dynamics that
is observed in the quantum case as $B,E > 0.3$ does not correspond
to any transition from mainly quantum to mainly classical
dynamics.

We now give a simple and intuitive explanation for the suppression of chaos
by the quantum correction, in our model.
Let us consider the
map of the plane $(\xi,\eta)$ obtained by a Poincar\'e section of the
Hamiltonian system defined by Eq.\ (\ref{H}), such as those reported in Figs.
\ref{fig_poincare_lowb} and \ref{fig_poincare_highb}.
As $B=0$ both the classical
and the quantum system are integrable. Hence the trajectories of the map
lie on invariant tori. Nevertheless, the geometry of such tori is dramatically
different in the two cases: in the classical one, as $E > 1/8$ a hyperbolic
fixed point at $X=(0,0)$ exists, and the trajectory that passes through
$X$ is actually a separatrix. Such a hyperbolic point is due to
the presence of a local maximum in the potential. The quantum correction
to the Hamiltonian is a ``centrifugal term'' that removes the local
maximum of the potential and replaces it
with an infinite barrier. Consequently,
no hyperbolic fixed point in $(0,0)$ exists in the quantum Poincar\'e section.
As soon as the perturbation $-B\varphi$ is turned on, chaos immediately
shows up in the classical case just in the neighborhood of $X$, because
the stable and unstable manifolds, which constituted the separatrix of
the unperturbed map, split and have infinite intersections \cite{Poincare}.
In the quantum case this major source of chaos is removed because no separatrix
exists in the unperturbed case, and chaos shows up only when the perturbation
has completely distorted the original shape of the potential. In
physical terms, the quantum suppression of classical chaos in
our model is due
to the fact that the quantum fluctuations are kept away from zero
--- consistently with Heisenberg's uncertainty principle --- by
a quantum term in the effective Hamiltonian.

To summarize, we have presented an example in which the phenomenon of quantum
smoothing of classical chaos not only clearly shows up, but also
finds a simple explanation on physical grounds. In order
to understand how general this explanation can be, further work is
needed. On one hand, the large $N$ expansion belongs to semiclassical
approximations, the validity of which, 
as far as chaos is concerned, can be questionable.
In fact, in some cases it has been explicitly found that the onset
of chaos is in correspondence to the breakdown of the approximation,
and that the exact evolution of the quantum expectation values is not sensitive
to the initial conditions \cite{elze,exact}.
Yet it has been argued that ``semiquantum chaos'' can be a real effect in
open quantum systems since they are driven in a semiclassical regime by
the interactions with the environment \cite{elze,zurek}.
On the other hand, our results on quantum and classical Lyapunov exponents
(see e.g. Fig.\ \ref{fig_lyap}) are 
in qualitative agreement with those reported in
Ref. \cite{Guglielmo}, where a quantum Lyapunov
exponent --- defined via the Bohm approach to quantum mechanics --- 
was found positive but smaller than the classical one in
a model of an hydrogen atom in an oscillating electric field. In Bohm's
theory \cite{bohm} particles obey classical equations of motion with
an additional force derived from a ``quantum potential'' that is
of order $\hbar^2$ as the quantum correction to the effective Hamiltonian
(\ref{H}) is. Bohm's equations of motion are exact, 
being completely equivalent to 
the standard quantum theory, but to write these equations 
for the system (\ref{ham}) would require the solution of the full
time-dependent Schr\"odinger equation, 
thus the analysis of the exact Bohmian dynamics of our model system is
practically unfeasible. 
Our treatment is approximate
but tractable, and the features it shares with Bohmian
mechanics are certainly suggestive.

We thank A. Barducci, G. Pettini and M. Pettini for
fruitful discussions and suggestions. LC acknowledges useful
discussions with P. Castiglione and C. Presilla.
This work has been carried out within the
EEC program Human Capital
and Mobility (contracts n. OFES 950200, UE ERBCHRXCT 94-0579).

\vspace{-1cm}

\begin{references}
\bibitem{normalization} The minimum of the classical potential energy
has been fixed to zero for each value of $B$.
\end{references}
\end{document}